# 3D Modeling of Absorption by Various Species for Hot Jupiter HD209458b

I. F. Shaikhislamov[1,3], M. L. Khodachenko[2], H. Lammer[2], A. G. Berezutsky[1],
I. B. Miroshnichenko[1], M. S. Rumenskikh[1]

*1) Institute of Laser Physics SB RAS, Novosibirsk, Russia*
*2) Space Research Institute, Austrian Academy of Sciences, Graz, Austria*
*3) Institute of Astronomy, Russian Academy of Sciences, Moscow, Russia*
*E-mail address: ildars@ngs.ru*

**ABSTRACT**

The absorption of stellar radiation observed by the HD209458b in resonant lines of OI and CII has not yet been satisfactorily modeled. In our previous 2D simulations we have shown that the hydrogen-dominated upper atmosphere of HD209458b, heated by XUV radiation, expands supersonically beyond the Roche lobe and drags the heavier species along with it. Assuming solar abundances, OI and CII particles accelerated by tidal forces to velocities up to 50 km/s should produce the absorption due to Doppler resonance mechanism at the level of 6-10%, consistent with the observations. Since the 2D geometry does not take into account the Coriolis force in the planet reference frame, the question remained to which extent the spiraling of the escaping planetary material and its actually achieved velocity may influence the conclusions made on the basis of 2D modeling. In the present paper we apply for the first time in the study of HD209458b a global 3D hydrodynamic multi-fluid model that self-consistently describes the formation and expansion of the escaping planetary wind, affected by the tidal and Coriolis forces, as well as by the surrounding stellar wind. The modeling results confirm our previous findings that the velocity and density of the planetary flow are sufficiently high to produce the absorption in HI, OI, and CII resonant lines at the level close to the in-transit observed values. The novel finding is that the matching of the absorption measured in MgII and SiIII lines requires at least 10 times lower abundances of these elements than the Solar system values.

## 1. INTRODUCTION

The study of exoplanets is one of the fast growing and intriguing fields in space science. Up to 10% of all exoplanets could be so called hot Jupiters (hereafter HJs) or warm Neptunes orbiting rather close to the host star, many of them at a distance shorter than 0.05 AU. *Lammer et al. (2003)* were the first to show that the hydrogen-dominated atmosphere of HJ is heated to several thousand Kelvin and driven to continuous dynamical expansion. One of the most studied transiting exoplanets is the HD209458b (*Charbonneau et al. 1999*). *Vidal-Madjar et al. (2003)* observed the primary transits of this planet in Lyα with the STIS spectrograph on board of HST and reported 15% absorption signal, located mostly in the high velocity blue wing of the line. Subsequent re-analyses of the same data led to slightly smaller (6-9%) and more symmetric absorption profiles (*Ben-Jaffel 2007, 2010; Vidal-Madjar et al. 2008*). This was attributed to the presence of a dense escaping planetary atmospheric material at the altitudes of up to 2–2.5 planetary radii.

Soon after the discovery of first HJs, 1D hydrodynamic simulations began to model the expanding planetary upper atmospheres, in order to describe the process of planetary mass loss and to interpret the measured transiting spectra (*Yelle 2004; García Muñoz 2007; Shaikhislamov at al. 2014,* and references therein). The results of these simulations showed that the extreme heating of the planetary upper atmosphere, due to absorption of the ionizing stellar XUV radiation, leads to a supersonic outflow of planetary gas in the form of an escaping planetary wind (hereafter PW), that overcomes the planetary gravitational binding and expands beyond the Roche lobe. All performed simulations agree regarding an average exospheric temperature of about $10^4$ K, an outflow velocity of about 10 km/s, and a mass-loss rate in the range $10^9$–$10^{12}$ g/s.

The first attempts to explain the Lyα transit observations of HD209458b were based on the mechanism of acceleration of planetary hydrogen atoms to high velocities by the stellar radiation pressure (*Vidal-Madjar et al. 2003, Lecavelier des Etangs et al. 2004, 2008*). However, the studies of *García Muñoz (2007), Ben-Jaffel (2007, 2008)*, and *Koskinen et al. (2010)*, based on 1D hydrodynamic models, led to the conclusion that the absorbing atomic hydrogen more likely lies within the Roche lobe, and the absorption over broad emission profile is caused by the natural line broadening. The same conclusion was also made by *Shaikhislamov at al. (2016), Khodachenko et al. (2017)* and *Shaikhislamov et al. 2018a*, who used a 2D multi-fluid simulations to treat self-consistently the generation of PW and its interaction with the stellar wind (hereafter SW) plasma.

Besides Lyα, the primary transit of HD209458b has been also observed with HST/STIS at other far-UV (FUV) wavelengths. In particular, *Vidal-Madjar et al. (2004)* reported absorption depths of 13±4.5% and 7.5±3.5% in OI ($2p^2$ 2P–$2p^2$ 2D) and CII ($2p^4$ 3P–$2p^4$ 3S) FUV resonance lines, respectively. Further analysis of the same data-set by *Ben-Jaffel & Hosseini (2010)* led to similar results (10.5±4.4% and 7.4±4.7% for OI and CII, respectively). According to *Linsky et al. (2010)*, the analysis of four additional FUV primary transit observations carried out with the COS spectrograph, on board of HST, confirmed the previously measured CII transit depth of 7.8±1.3%. *Linsky et al. (2010)* reported also the detection of 8.2±1.4% absorption signal at the position of SiIII (1206.5 Å) resonance line in contrast with Vi*dal-Madjar et al. (2004)* who instead reported a non-detection. It should be noted that in view of possible strong stellar variability, *Ballester & Ben-Jaffel (2015)* called into question the detections of CII and SiIII absorption features with COS. The latest up to date observations of HD209458b carried out with HST/STIS, reported by *Vidal-Madjar et al. (2013)*, showed 6.2±2.9% absorption in the MgI resonance line, while no absorption was detected in the MgII h&k lines. This is a surprising result, because MgII absorption has been previously detected for another hot Jupiter, WASP-12b (*Fossati et al. 2010; Haswell et al. 2012*), and Mg is known to be rapidly ionized by the stellar FUV flux. All these observational results suggest the presence of heavy species in the upper atmosphere of HD209458b.

Previous works based on 1D hydrodynamic models were not able to find reasonable conditions within which the simulated OI, CII, and SiIII absorptions are comparable to (or even exceed) those observed in Lyα. For example, *Ben-Jaffel & Sona Hosseini (2010)*, using the density distributions provided by *García Muñoz (2007)*, obtained the absorption signatures of just 3.9% and 3.3% for the OI and CII multiplets, respectively. A number of processes have been suggested to account for this discrepancy. *Ben-Jaffel & Sona Hosseini* (*2010*) suggested preferential heating of minor heavier species inside the Roche lobe to explain the absorption and concluded that the effective temperatures should be about 10 times larger for oxygen and 5 times larger for carbon particles than for hydrogen. However, the existence of such a large temperature difference is hardly possible for the entire population of heavier species, mostly because the hydrogen density inside the Roche lobe is higher than $10^7$ cm$^{-3}$ and the partially ionized plasma is strongly collisional. *Koskinen et al.* (*2010*) calculated the absorption by OI atoms based on a 1D hydrostatic model, obtaining an absorption signature of 4.3%. A number of other factors, which could increase the OI absorption, were further investigated by *Koskinen et al.* (*2010*), including sporadic hot spots on the stellar disk, limb darkening or limb brightening, super-solar abundances, and the position of the $H_2$/H dissociation front. However, none of the considered effects could provide an increase of the OI transit depth, comparable to, or higher, than that of HI. Follow up computations, based on the same model, led to the absorption of 5.2% and 4.6% for CII and SiIII, respectively (*Koskinen et al. 2013b*).

The crucial point here is that, while Lyα absorption can be explained by natural line broadening with a sufficiently large number of hydrogen particles inside the Roche lobe, the density of heavier species is too small for that, whereas the typical temperature of $10^4$ K is too low to produce sufficient resonant absorption due to the thermal line broadening over the typical line half-widths of 20–30 km/s. At the same time, the necessary broadening may be generated by a global flow of planetary material outside the Roche lobe, as it was qualitatively proposed by *Tian et al.* (*2005*). To calculate Doppler line broadening absorption by accelerated particles beyond the Roche lobe at least a 2D modeling is needed. This modeling and the related study of absorption features were done for HD209458b for the first time in our previous paper (*Shaikhislamov et al. 2018a*).

Using 2D axisymmetric geometry in a fully self-consistent multi-fluid numerical model (*Shaikhislamov et al. 2014, 2016, Khodachenko et al. 2015, 2017*), *Shaikhislamov et al. 2018a* have shown that HD209458b transit depths at a level of about 10% observed in OI and CII resonant multiplets can be reproduced assuming typical for Solar system abundances of oxygen and carbon. The performed simulations of the escaping PW, interacting with the background SW, confirmed that the upper atmospheric material expanded to the

Roche lobe and accelerated beyond it by tidal forces to velocities of several tens of km/s, would drag the heavy elements due to collisional coupling. This sufficiently increases the absorption of stellar OI and CII lines due to Doppler resonant absorption mechanism, providing values close to the observed ones. However, calculation of resonant absorption by SiIII and MgI, MgII lines under the same conditions have shown that to match the observations, the abundances of these species in the upper atmosphere of HD209458b should be much less than the typical Solar System values.

Nevertheless, the applicability of 2D axisymmetric geometry to the modeling of planetary atmospheric material outflow is restricted to a relatively close region around the planet. At distances larger than $(5–10)R_p$, the streams of escaping PW driven by the tidal force would bend clockwise due to the Coriolis force, resulting in decrease of acceleration and the flow velocity component along the star-planet line. This effect follows from the momentum conservation law, which prevents the direct fall of orbiting material on a star, forcing instead its spiraling. Thus, the increase of the PW velocity with distance along the tidally pulled streams, obtained in the 2D axisymmetric modeling by *Shaikhislamov et al. (2018a)*, is most likely an overestimation. The present work remedies this deficiency by using the fully 3D simulation.

The results of 3D multi-fluid self-consistent modeling reported here show significantly more complex spatial structure of the escaping upper atmospheric material flow, and slower PW velocities, as those obtained in 2D simulations. However, previous 2D modeling results regarding the absorption still remain valid in most details, though there are some quantitative differences. The main finding here is that the escaping PW of HD209458b, simulated at the supposed parameters, is sufficiently fast to generate the observed significant Doppler resonant absorption in resonant lines of OI and CII, assuming the Solar abundancies of those elements. At the same time, to match the measured absorption in the resonant lines of MgII and SiIII, the magnesium and silicon abundances should be at least 10 times lower, than the corresponding Solar system values.

The paper is organized as follows. In Section 2, we briefly describe the model and its novel features. In Section 3.1 the absorption in HI, OI, CII, SiIII, MgI, and MgII lines is calculated for HD209458b under the conditions of a weak SW (i.e. low stellar mass loss rate). In Section 3.2 we explore at which parameters of SW its interaction with the PW affects the absorption in lines of interest. Section 4 presents the discussion and conclusions.

## 2 THE NUMERICAL MODEL AND ABSORPTION PARAMETERS

The modeled HD209458 system consists of a planet with mass $M_p=0.71M_J$ and radius $R_p=1.38R_J$ orbiting at a distance of $D=0.047$ a.u. around a solar-type *G*-star with the mass $M_{st}=1.148M_{Sun}$ and radius $R_{st}=1.13R_{Sun}$. The planetary and stellar radii and masses are given in the units of of the Jupiter and the Sun, respectively. As the characteristic values of the numerical problem, we use like in our previous simulations (*Shaikhislamov et al. 2014, 2016, Khodachenko et al. 2015, 2017*), the planet radius $R_p$, the temperature $10^4$ K, and the corresponding thermal velocity of protons $V_o=9.07$ km/s. The obtained numerical solutions are correspondingly scaled in these units.

The 3D multi-fluid hydrodynamic model used in the present work has been already described in (*Shaikhislamov et al. 2018b, Khodachenko et al. 2019*). It was developed on the base of previously employed 2D code (*Khodachenko et al. 2015, 2017, Shaikhislamov et al. 2016*). Here we just repeat the most important details regarding the model. The code solves numerically the hydrodynamic equations of continuity, momentum, and energy for all species of the simulated multi-component flow. Among the considered fluid species are the hydrogen and helium particles (H, $H^+$, He, $He^+$) and the heavier species: O, C, Mg, Si. While the model in its general version includes also the complex hydrogen species, such as $H_2$, $H_2^+$, and $H_3^+$ (*Shaikhislamov et al. 2018b, Khodachenko et al. 2019*), in the present study we disregard them. The reason for such simplification in the case of HD209458b is their dissociation at relatively low altitudes. As it was shown in the detailed 1D aeronomy simulation of HD209458b by *García Muñoz (2007)*, the presence of oxygen with the abundancies above 0.01 of the Solar system value leads to a rapid decay of the $H_2$ population with height leaving only the hydrogen atoms and protons. Without account of hydrogen molecules, the H and He chemistry is governed by photo-ionization, dielectronic recombination, electron impact excitation and ionization. Photo-ionization also results in the strong heating by the produced photo-electrons, which drives the hydrodynamic outflow of the planetary atmospheric material. The corresponding heating term is derived by integration of the stellar XUV spectrum, taking into account the attenuation of photons in the atmosphere according to the wavelength-dependent cross-section. As a proxy, we use here the solar spectrum, assuming slightly larger integrated XUV flux (λ<91.2 nm) of about $F_{XUV}=10$ erg s$^{-1}$ cm$^{-2}$ at a reference distance of 1 a.u., reflecting the slightly larger size of HD209458.

In the present work we reconsider the heating efficiency by photo-electrons. In previous modeling we used an overall efficiency of 0.5 assuming that half of the total energy of photo-electrons generated by $F_{XUV}$ flux is lost via excitation of hydrogen and emission of Lyα photons which are reabsorbed, reemitted and eventually lost (*Shaikhislamov et al. 2014*). Other channels which keep the energy in the system are further ionization of particles by electron impact and elastic collisions. However, the helium, if present, changes the channels of energy loss. First, to create photoelectron with an energy $E_{el}$ sufficient for the excitation of hydrogen atoms ($I_{H,exc}=10.2$ eV), the primary photon should have energy $E_{ph}$ close to the He ionization potential ($I_{He}=24.6$ eV). This means that neutral helium would capture such primary photons and, therefore, decrease the emission of Lyα radiation. Since the cross-section of He photo-ionization at $E_{ph}>24.6$ eV is an order of magnitude higher than that of the HI photo-ionization, this effect is not small. Second, even at relatively low abundance ~3% helium significantly increases, due to its photo-ionization, the density of electrons in the upper atmosphere below the thermosphere maximum (*Shaikhislamov et al. 2018b*). Therefore, the photoelectrons would lose their energy via Coulomb collisions more rapidly. Using the typical density profiles of HD209458b upper atmosphere derived in our simulations we plotted in Figure 1 the rates of energy loss of photo-electron via Coulomb collisions with background electrons (*Braginskii 1965*) and via excitation of hydrogen (*Dalgarno & McCray 1972*). One can see that Coulomb collisions dominate everywhere except the deep upper atmosphere ($r<1.3R_p$, $n_H>10^{10}$ cm$^{-3}$) where density of atomic hydrogen steeply increases, and the attenuation of XUV flux cuts off the photo-ionization. In the present paper, we calculate the heating more accurately by integrating the XUV spectra and taking into account all the above factors. Specifically, at $E_{el}<10.2$ eV all the photo-electron energy (either $E_{el}=E_{ph}-I_{H,exc}$ or $E_{el}=E_{ph}-I_{He}$ depending on ionization of H or He) goes to heating. At $E_{el}>10.2$ eV we calculate how much energy is lost via Lyα emission $\Delta E=E_{el}-I_{H,exc}$ by weighting rates of HI excitation, ionization and collisions with background electrons. It has been found that the obtained results are practically identical with those obtained assuming constant heating efficiency, equal to 1, rather than 0.5. Note that the net efficiency, defined as the ratio of heating to the total XUV flux absorbed, is actually quite small, ≈17%, because the most of energy is spent for the ionization.

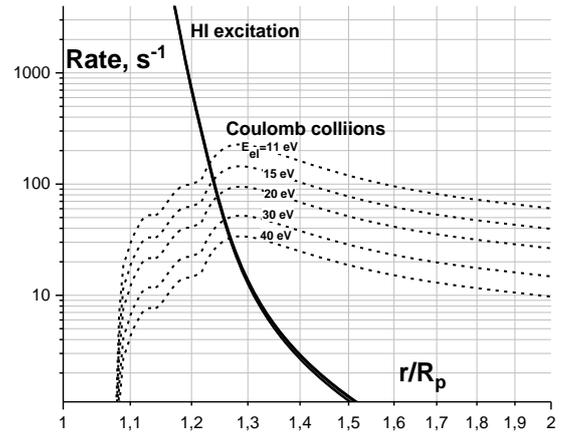

**Figure 1.** Photo-electron energy loss rates at different heights in the upper atmosphere of HD209458b via Coulomb collisions with the background electrons (dashed lines) and via excitation of hydrogen atoms (solid lines). Test photo-electrons with different energies are plotted, as indicated. The atmospheric parameters (electron density and temperature, hydrogen density) were obtained in one of the typical simulation runs discussed below.

The model equations are solved in a non-inertial spherical frame of reference fixed at the planet center, which orbits together with the planet and rotates with the same rate, so that it is continuously faces the star with the same side. The polar-axis Z is directed perpendicular to the ecliptic plane. This is a so-called tidally locked spherical frame of reference. In this frame we properly account the non-inertial terms, i.e., the generalized gravity potential and Coriolis force. The fluid velocity at the planetary surface is taken to be zero. To keep the number of points in the numerical code tractable for processing, the radial mesh is highly non-uniform, with the grid step increasing exponentially from the planet surface. This allows resolving of the highly stratified upper atmosphere of the planet, where the required grid step is as small as $\Delta r=R_p/400$. Taking into account that the planet-star distance is about $70R_p$, the corresponding grid step of ~$R_p/10$, at the distances of several $R_p$, ensures sufficient resolution of the model also in these remote regions. The exponential radial spacing in the spherical coordinate system also keeps the same resolution in all 3 dimensions, if the azimuthal and latitudinal mesh is chosen so that $\Delta\varphi\approx\Delta\theta\approx\Delta r/r$. For the analysis of simulation results, a related Cartesian coordinate system with X-axis pointing from the planet to the star and Z-axis perpendicular to the ecliptic plane is also used.

For the self-consistent calculations on the scale of the whole star-planet system, we incorporate in the same code the SW plasma dynamics as well.

Besides the upper planetary atmosphere, the stellar corona is another boundary of the simulation domain, at which the corresponding coronal values of the plasma parameters are fixed. Outside this boundary, taken for simplicity at the stellar surface with the radius $R_{star}$, the SW, i.e. the proton fluid, is calculated by the same code as one used for the protons of PW. The simulation of SW in the acceleration zone $<10 R_{star}$ is a complex problem, which has been extensively studied in the last decades (see e.g., *Usmanov et al. 2011* and references therein). Its detailed treatment, especially the processes of SW heating and acceleration up to super-sonic velocities (still not yet fully understood), are beyond the scope of the present paper. A common approach of the majority of the global astrophysical codes, adopted for the simulation of exoplanetary environments, is to model SW in a simplified way, using a polytropic or even an isothermal specific heat ratio $1 \leq \gamma_p \leq 1.2$ (e.g., in *Bisikalo et al. 2013, Matsakos at al 2015, Christie et al. 2016, Daley-Yates & Stevens 2018*). Any polytropic index, lower than the adiabatic one $\gamma_a=5/3$, introduces in the energy equation an effective source term which replaces the variety of complex SW energizing processes. However, simulation with a non-adiabatic index makes unrealistic the treatment of shocks, which develope in the region of the colliding SW and PW flows. To avoid this difficulty, we introduce in energy equation an empirical heating source to drive the SW. The corresponding source term is found from a semi-analytical solution of 1D polytropic Parker-like model, which yields for a given stellar gravity, the base temperature $T_{cor}$, and the asymptotic supersonic velocity $V_{sw,\infty}$, a unique value of $\gamma_p$ and the radial profiles of all SW parameters (see, e.g., in *Keppens & Goedbloed 1999*). To fix the SW plasma density at the stellar surface boundary, we use an integral characteristic parameter – stellar mass loss rate, $M'_{sw}$. The heating term, therefore, is found as:

$$W_{sw}(r) = (\gamma_a - \gamma_p) \cdot T_p(r) \cdot \mathrm{div} V_p(r) \qquad (2.1)$$

Here, $T_p$ and $V_p$ are obtained from the polytropic solution, and the distance $r$ is measured from the center of star. It was specially checked in the simulations that without PW, the solution obtained for SW is in a good agreement with the analytical model. Note, that the equation (2.1) specifies a spatially distributed, continuous and stationary in time heat source, which is defined by the specified above parameters only. To avoid of unphysical artefacts, the heating source is switched off in the regions where a significant amount of particles of the planetary origin penetrate into the SW. The modelling approach, described above, allows a physically self-consistent global simulation of the generation of super-sonic PW and SW, as well as their interaction.

As the initial state, we consider a fully neutral atmosphere in barometric equilibrium. At the inner boundary of the simulation domain $r=R_p$ we fix the temperature in the range of $(1-2) \cdot 10^3$ K, the pressure $p=0.05$ bar and the mixing ratio of elements in relation to atomic hydrogen. The minor species are added to the initial atmosphere assuming, as a first approximation, the Solar system abundances (*Anders & Grevesse 1989*). In this paper we treat the ionization stages of O, C, Mg, and Si more accurately than in previous works (*García Muñoz 2007, Koskinen et al. 2013a, Shaikhislamov et al. 2018a*). The population of different elements includes three ionic stages which are calculated assuming the specific photo-ionization (*Verner et al. 1996*) and recombination rates (*Le Teuff et al. 2000, Nahar & Pradhan 1997*). For the O, C and Mg we take the stages from the atomic to double-ionized ones. Double-ionized ions of these species have large ionization potentials and very large photo-ionization times. Therefore, at considered conditions they can be treated as terminal states. For the Si we calculate the stages of SiII, SiIII and SiIV. The production of SiV can be safely neglected. We checked that including in simulations the SiI stage doesn't change the abundancies of SiII and SiIII ions, because SiI is very rapidly photo-ionized in the upper atmosphere (*Koskinen et al. 2013a*). Like in previous works, we take into account that the populations of OI and OII, as well as SiII and SiIII, are strongly affected by the resonant charge-exchange reaction with the atomic hydrogen and protons: $O + H^+ \leftrightarrow O^+ + H$, $Si^+ + H^+ \leftrightarrow Si^{2+} + H$. We found out in tests that neglecting OIII, CIII, MgIII and SiIV ions in the production equations results in a factor of 20% change in densities of elements.

As it has been pointed out in our works (*Shaikhislamov et al. 2016, 2018a, Khodachenko et al. 2015, 2017*), for the typical parameters of plasmaspheres of the close-orbit hot exoplanets, the Coulomb collisions with protons effectively couple all ions. Therefore, there is no need to calculate the dynamics of every charged component of the plasma fluid species, and we assume in the simulations all of them (i.e., $H^+$, $He^+$, $O^{n+}$, $C^{n+}$, $Mg^{n+}$, $Si^{n+}$) to have the same temperature and velocity. On the other hand, the temperature and velocity of each neutral component are calculated individually. Nevertheless, the neutral atoms usually are more or less coupled to the main flow as well, due to elastic collisions and charge exchange (*Shaikhislamov et al. 2016, 2018a; Khodachenko et al. 2017*).

The spectral absorption is described by the so called Voigt convolution of a Lorentz line shape with a natural width and Maxwellian distribution, characterized by the mean temperature $T$, particle density $n$, and component of velocity $V_x$ along the line of sight (LOS). We use an empirical analytic fit of the Voigt integral which has the accuracy better than 1% at temperatures >2 K (*Tasitsiomi 2006*). The whole set of used formulae is given in *Shaikhislamov et al. (2018a)* and *Khoduchenko et al. (2017, 2019)*.

It is worth to mention, that the applied analytic fit of the Voigt integral explicitly shows that the absorption consists of two components. Of which the first one is the resonant absorption by atoms matching the Doppler shifted velocity of the line profile. This process is characterized by a significant cross-section within the line core. When the average thermal velocity of absorbing particles is comparable to the corresponding velocity $V$ of particular part of the spectral line, then the number of resonant atoms with $|V_x|\pm(2kT/m)^{1/2}\approx V$ is statistically large due to the width of Maxwell distribution function. In this case the produced absorption is related to the thermal line broadening. The heavier particles at the same temperature have smaller thermal dispersion of the distribution function, and therefore the smaller velocity range for the thermal line broadening. That is the particular reason why it is difficult to explain the absorption in O, C lines only by thermal line broadening. Besides of that, a similar resonant absorption will take place when a relatively high bulk velocity of particles motion appears in Doppler resonance with a given part of the line profile. This produces absorption which is particularly relevant for the present study. Further on for simplicity's sake, the absorption produced due to the matching between the flow bulk velocity and the spectral line velocity range we call *Doppler resonance absorption*. The second component of the total absorption is due to the non-resonant process related to the natural line broadening affecting the far wings of the Lorentzian profile, which has a much smaller cross-section.

Table 1 summarizes the properties and relevant parameters of the spectral lines considered in this paper, specifically, the width of line, the measured absorption, the statistical weights of the ground states and the relative intensities measured out of transit for each line in multiplet, the absorption cross-sections: one due to the resonant Doppler broadening ($\sigma_{res}$) and another due to natural line broadening ($\sigma_{nat}$) mechanism. The intensity of unresolved transition without orbital momentum change (3/2–3/2) in the CII multiplet is taken to be zero, because its oscillator strength is an order of magnitude lower, as compared to that of the nearby transition 3/2–5/2. This fact results in double decrease of the overall absorption of the multiplet. The CII and SiIII lines are sufficiently well resolved in the observed spectra (*Linsky et al. 2010*). For these lines we employ the same analytical fits as those used in *Koskinen et al. (2013b)* and *Shaikhislamov et al. (2018a)*. For OI, the intrinsic stellar emission line shapes are taken from the solar spectrum. This assumption is justified by the strong similarity between HD209458 and the Sun (*Shkolnik et al. 2005*). The double peaked lines of the OI multiplet were analytically approximated by a double Gaussian profile, with the corresponding parameters from *Link et al. (1988)*.

The used shapes for CII, OI, and SiIII lines are shown in Figure 2. In particular, the OI lines have half-width of about 22 km/s, while that of CII and SiIII lines about 35 km/s. The line cores in the range ±7 km/s are not considered, because they are affected by the interstellar medium absorption. Note, that this reduces the overall absorption of OI and CII lines.

**Table 1.** Parameters of the considered spectral lines and absorption cross-sections.

| El. | Line Å | Width ±ΔV km/s | Measured. absorption % | Statw. /Int. | $\sigma_{res}$ cm² *7 | $\sigma_{nat}$ cm² *8 |
|---|---|---|---|---|---|---|
| OI | 1302.2 | 22 | 10.5±4.4 *1 | 5/1 | 2.9·10⁻¹⁴ | 3.3·10⁻¹⁷ |
|  | 1304.9 | 22 |  | 3/1.5 | 2.9·10⁻¹⁴ | 2.0·10⁻¹⁷ |
|  | 1306.0 | 18 |  | 1/1.2 | 2.9·10⁻¹⁴ | 6.5·10⁻¹⁸ |
| CII | 1334.5 | 24 | 7.4±4.7 *2 | 2/0.7 | 6.9·10⁻¹⁴ | 5.4·10⁻¹⁷ |
|  | 1335.7 |  |  | 4/0 | 6.9·10⁻¹⁵ | 1.1·10⁻¹⁸ |
|  | 1335.7 | 35 | 7.8±1.3 *3 | 4/1 | 6.2·10⁻¹⁴ | 5.8·10⁻¹⁷ |
| SiIII | 1206.5 | 33 | 8.2±1.4 *3 | 1/1 | 1.3·10⁻¹² | 1.4·10⁻¹⁴ |
| MgI | 2853.0 | 400 | 6.2±2.9 *4 | 1/1 | 3.0·10⁻¹² | 4.8·10⁻¹⁶ |
| MgII | 2795.5 | 400 | < 2 | 1/1 | 9.5·10⁻¹³ | 8.0·10⁻¹⁷ |
| HI | 1215.6 | 80 | 8.6±2.0 *5  6.6±2.2 *6 | 1/1 | 5.9·10⁻¹⁴ | 2.6·10⁻¹⁷ |

*1 in the interval 1299-1310 Å (*Vidal-Majar et al. 2004*).
*2 in the interval 1332-1340 Å (*Vidal-Majar et al. 2004*).
*3 in the interval ±50 km/s (*Linsky et al. 2010*).
*4 in the interval −62−19 km/s (*Vidal-Majar et al. 2013*).
*5 in the interval ±200 km/s, medium resolution data (*Ben-Jaffel 2007, 2008*).
*6 in the interval ±1000 km/s, low resolution data (*Ben-Jaffel & Sona Hosseini 2010*).
*7 cross-section calculated at T=10⁴ K.
*8 cross-section calculated at V=10 km/s.

For the Lyα line we use an out-of-transit profile measured by *Vidal-Majar et al. 2003*. The interval ±50 km/s is excluded from the consideration because of geocoronal contamination. Magnesium lines are rather different from the CII, OI, and SiIII lines, because of their very large width and uncertainty in the measured absorptions. As an analytical proxy of their emission profiles, which have complex inverted shape, we used the measurement data from *Vidal-Majar et al. 2013*.

For comparison with observations we average simulated absorption over the interval ±0.5 Å for CII, OI, SiIII lines, ±4 Å for MgI, MgII lines and ±1 Å for Lyα line.

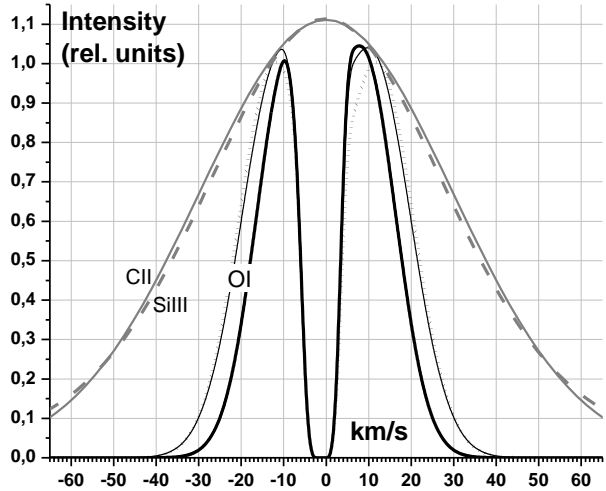

**Figure 2.** Analytical fit of line shape for the OI multiplet (1302.17 Å – dotted, 1304.86 Å – thin solid, 1306.02 Å – thick solid), CII line 1335.71 Å (gray solid) and SiIII line 1206.5 Å (gray dashed).

## 3 SIMULATION RESULTS

### 3.1 Absorption under a very week SW

As will be shown later, the interaction with SW plasma does not influence significantly the absorption in lines of minor elements, unless SW is very strong. Therefore, we begin with a simple case of a very weak SW, when the stellar mass loss rate is two orders of magnitude less than that of the Sun ($\approx 2.5 \cdot 10^{12}$ g/s), and study the influence of other parameters important for the planetary mass loss. These are the stellar XUV flux $F_{XUV}$, helium and other elements' abundances, as well as the base temperature of the upper atmosphere $T_{base}$. The corresponding material escape in this regime appears to be close to one, we called as 'captured by the star' in our previous 2D simulations (*Shaikhislamov et al. 2016, 2018a, Khodachenko et al. 2017*).

Figure 3 shows the color plots of density distribution of HI, HII, OI and CII in the escaping PW of HD209458b on the whole system scale under the conditions which we consider as the *standard* ones: $F_{XUV}$=10 erg cm$^{-2}$ s$^{-1}$ at a reference distance of 1 a.u., Solar system abundancies of elements (He/H=0.05, O/H=8.5·10$^{-4}$, C/H=3.6·10$^{-4}$, Si/H=3.7·10$^{-5}$, Mg/H=3.7·10$^{-5}$), and $T_{base}$=1500 K. This set of the modeling parameters is listed under the number N1 in Table 2, which summarizes all simulations discussed in the paper.

**Table 2.** List of simulation scenarios, corresponding modeling parameters, and the results of calculations. Columns from left to right: number of simulation run, the assumed value of stellar XUV flux $F_{XUV}$ in erg cm$^{-2}$ s$^{-1}$ at 1 a.u., calculated mass loss rate by the planet in units of $10^{10}$ g/s; calculated absorption in lines of the corresponding elements in % (details on the line parameters are given in Table 1). The last column indicates additional modeling conditions (if any) applied in the particular simulation runs, e.g., different helium abundances at the base atmosphere of the planet (rows 2–4), base temperature in K (rows 8–10), the SW mass loss in $10^{12}$ g/s (rows 10–12);

| N | XUV | M′ | HI | OI | CII | SiIII | MgI | MgII | * |
|---|-----|-----|-----|-----|-----|-------|-----|------|---|
| 1 | 10 | 27 | 7.6 | 7.6 | 6.8 | 38 | 2.1 | 5.6 | |
| 2 | 10 | 31 | 9.9 | 9.4 | 7.5 | 31 | 2.2 | 6.8 | He/H 0.01 |
| 3 | 10 | 28 | 6.0 | 5.7 | 6.6 | 44 | 2.1 | 4.3 | 0.1 |
| 4 | 10 | 24 | 4.6 | 3.6 | 5.4 | 48 | 2.0 | 3.0 | 0.2 |
| 5 | **5** | 16 | 6.8 | 7.2 | 7.3 | 37 | 2.2 | 3.6 | |
| 6 | **20** | 50 | 8.9 | 6.7 | 6.4 | 36 | 2.2 | 7.6 | |
| 7 | **30** | 69 | 9.3 | 6.0 | 6.3 | 26 | 2.2 | 8.5 | |
| 8 | 10 | 24 | 6.1 | 6.1 | 5.9 | 36 | 2.0 | 5.0 | $T_{base}$ 10$^3$ K |
| 9 | 10 | 35 | 9.6 | 9.7 | 7.9 | 40 | 2.4 | 6.5 | 2·10$^3$ |
| 10 | 10 | 27 | 7.8 | 7.6 | 7.0 | 38 | 2.1 | 5.3 | M′$_{SW}$ 2.5 *1 |
| 11 | 10 | 27 | 8.0 | 7.3 | 8.1 | 38 | 2.1 | 4.5 | 7.5 *2 |
| 12 | 10 | 27 | 18 | 5.4 | 5.2 | 13 | 2.1 | 3.2 | 25 *3 |
| 13 | 10 | 8 | 3.5 | 2.5 | 5.0 | 15 | 2.5 | 5.0 | H$_2$ *4 |
| 14 | 10 | 16 | 4.3 | 4.1 | 13 | 33 | 1.9 | 3.6 | H$_2$ *5 |

*1 $n_{sw}$=10$^4$ cm$^{-3}$, $V_{sw}$=220 km/s, $T_{sw}$=0.8 MK, at the planet orbit.
*2 $n_{sw}$=3·10$^4$ cm$^{-3}$ ($V_{sw}$, $T_{sw}$ the same as in case *1).
*3 $n_{sw}$=10$^5$ cm$^{-3}$ ($V_{sw}$, $T_{sw}$ the same as in case *1).
*4 Simulation with full hydrogen chemistry, heating efficiency of 0.5, and weak stellar wind.
*5 Simulation with full hydrogen chemistry, heating efficiency of 1.0, and weak stellar wind.

From the HI and HII density distribution plots one can see that under the conditions of a very weak, i.e., negligible, SW the escaping PW material of HD209458b forms a kind of an accretion torus around the star (*Debrecht et al. 2018, Khodachenko et al. 2019*). Its width in the equatorial plane is about 40$R_p$. In the meridional plane the torus size is restricted by the stellar gravity action and reaches values of about 20$R_p$. The detailed study of the planetary material accretion process is beyond the scope of the present paper. It is worth however to note, that due to the photoionization of species, the accretion torus does not influence the absorption, which mainly takes place within the range of distances <$10R_p$ around the planet. In particular, as shown in Figure 3, the HI population, because of the photo-ionization, does not extend far from the planet. The distributions of minor species CII and OI have even smaller scales. They are shown in Figure 3 with a two- and a four- times zoom, respectively. Due to the photo-ionization, the region of relatively high density of OI and CII is also restricted within the range of about $10R_p$ around the planet. In the equatorial plane, the structure of PW flow is strongly distorted by the Coriolis force. For more detailed quantitative representation, the density profiles of absorbing species of interest along Z-axis are given in Figures 4 and 5. It can be seen that the half-ionization points for hydrogen and oxygen are at about ≈$3R_p$, for carbon at ≈$1.5R_p$, and for magnesium at ≈$1.2R_p$. There is a sharp boundary at the distance of about $10R_p$ above and below the equatorial plane, beyond which the density of elements sharply drops. This is the effect of the stellar gravity (*Trammel et al. 2014, Shaikhislamov et al. 2018a*). The maximum temperature of thermosphere is slightly higher than $10^4$ K and the escaping PW velocity reaches ~20 km/s. The gradual increase of temperature and velocity beyond $10R_p$ is due to the presence of rarefied stellar plasma.

**Figure 3.** The density distribution of species (in log-scale) obtained in the simulation run N1. From top to bottom: hydrogen atoms ($n_{HI}$), carbon ions ($n_{CII}$), protons ($n_{HII}$), oxygen atoms ($n_{OI}$). The values outside the indicated variation ranges are colored either in red if smaller than minimum, or in blue, if higher than maximum. The first two panels show the equatorial cuts, and the bottom two the meridional cuts, respectively. The distance in each plot is scaled in units of $R_p$. Black circles indicate the star or the planet. Black lines show the streamlines of corresponding elements.

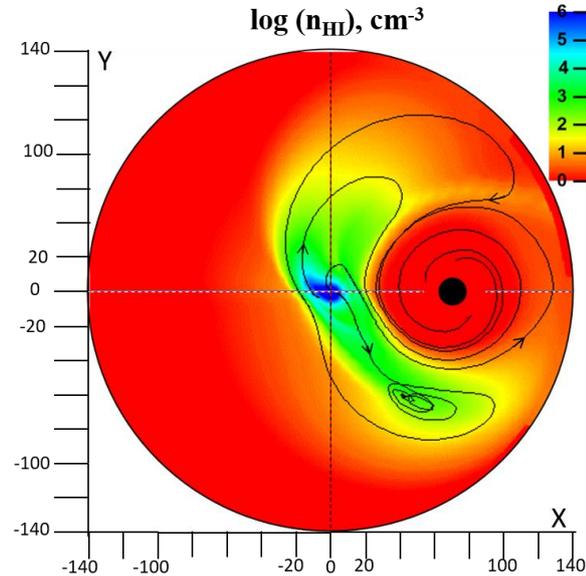

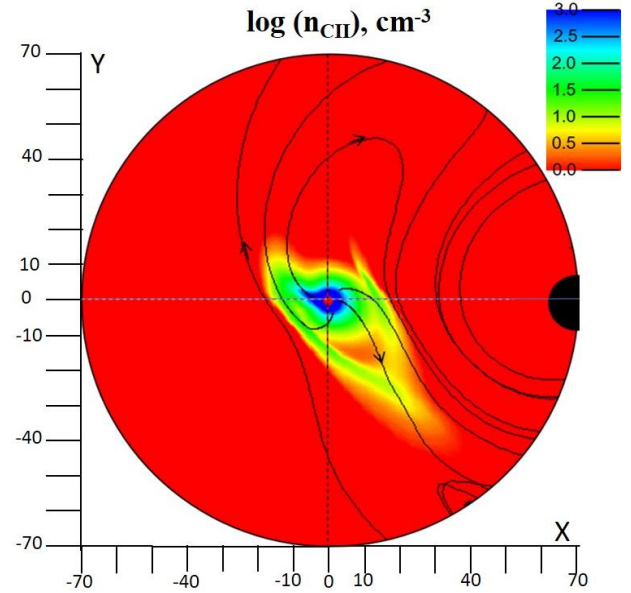

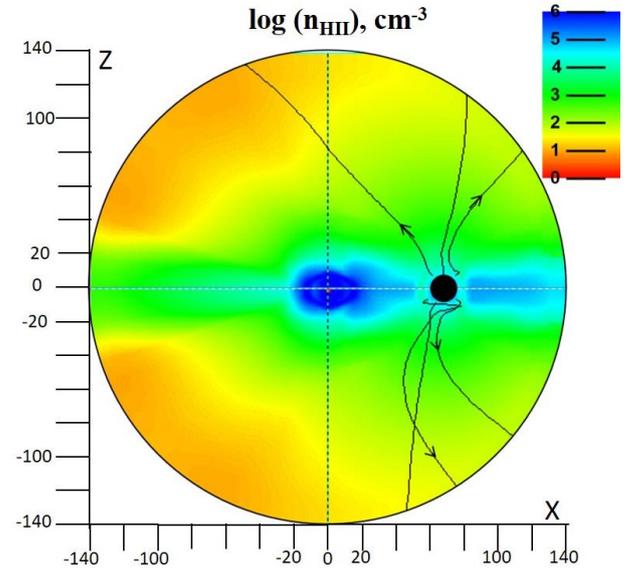

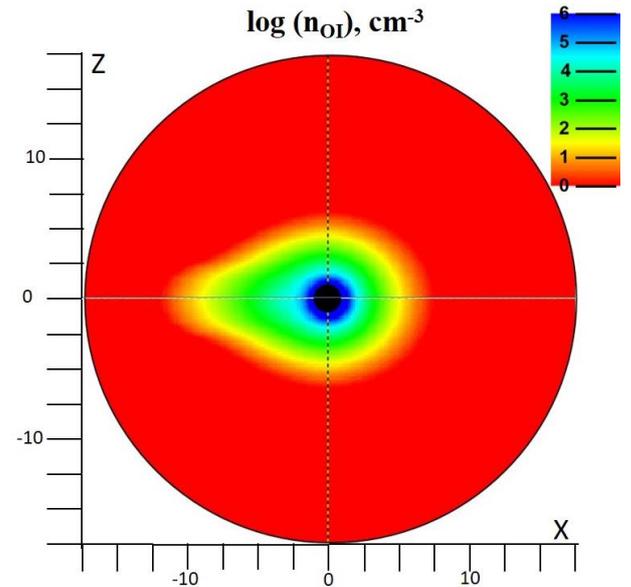

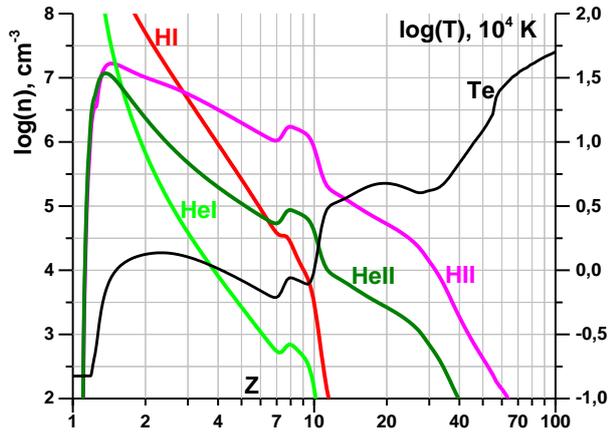

**Figure 4.** The density profiles (*left axis,* log-scale) of hydrogen atoms ($n_{HI}$), protons ($n_{HII}$), neutral helium ($n_{HeI}$), and ionized helium ($n_{HeII}$) along Z-axis (scaled in units of $R_P$) obtained in the simulation run N1. Black line shows the temperature profile (*right axis,* log-scale).

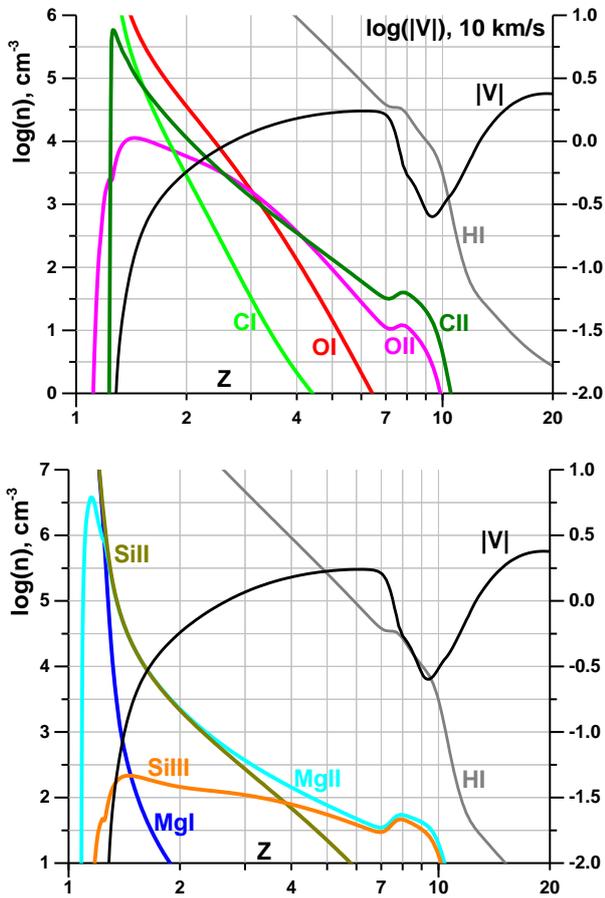

**Figure 5.** The density profiles (*left axis,* log-scale) of minor species and their ions along Z-axis obtained in the simulation run N1. *Top panel:* oxygen and carbon; *Bottom panel:* magnesium and silicon ions. For comparison, the atomic hydrogen density profile is shown in both panels ($n_{HI}$, gray line). Black line shows the profile of absolute value of the PW flow velocity (*right axis,* log-scale).

A crucial feature for the present study is the value of the LOS velocity $V_x$ of absorbing agents, dragged by the escaping PW, obtained in the 3D geometry with the correct simulation of Coriolis force and tidal effects. It is illustrated in Figure 6 with a set of velocity profiles along X-axis at Z=0 and different offset distances along Y. One can see that the LOS velocity does not exceed 30 km/s in the direction from the star and 40 km/s towards the star. This is significantly less than the values predicted in 2D simulations, but well coincides with the typical widths of the considered spectral lines (see table 1). Therefore, corresponding absorption can be expected to remain at about the same level as in *Shaikhislamov et al. 2018a*. As summarized in Table 2, the absorption for OI triplet is about the same as in 2D simulations, i.e. 7.6%, while for CII triplet it is 2% less, which is mainly due to more accurate calculations of the carbon ionization stages. The out-of-transit and in-transit profiles of HI line and the strongest lines in OI and CII triplets are shown in Figure 7. One can see that the absorption is generally symmetric, except of OI line which shows a higher level in the blue wing of the line. For heavy species such as OI and CII, the most of absorption is at the Doppler shifted velocities 10–30 km/s. At higher velocities, the absorption rapidly decreases due to decreasing number of absorbing particles with $|V|$>30 km/s. Note, that this absorption comes from the Doppler resonance broadening and is produced by particles which are dragged by the escaping PW flow and which have velocities matching the spectral line profile. At the same time, the absorption produced only by thermal Doppler broadening due to Maxwell distribution of particles, calculated for comparison for CII line, assuming zero bulk velocity (dashed line in the bottom panel in Figure 7), appears to be rather small.

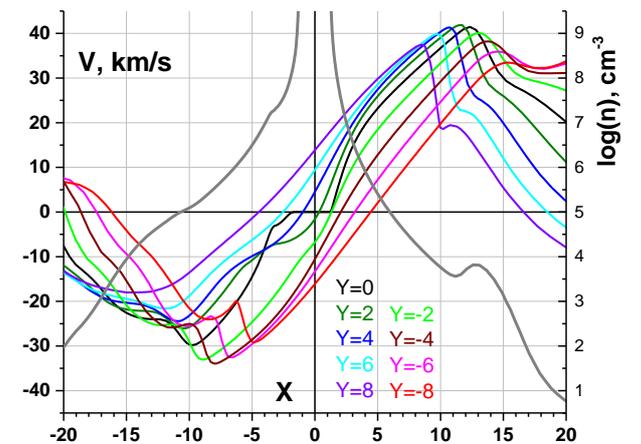

**Figure 6.** Profiles of the LOS velocity component $V_x$ (*left axis*) of the PW flow versus X coordinate, obtained in the simulation run N1, for Z=0 and different values of Y coordinate (indicated with different colors). To compliment the picture, gray line shows the density profile of HI (at Y=Z=0, *right axis*).

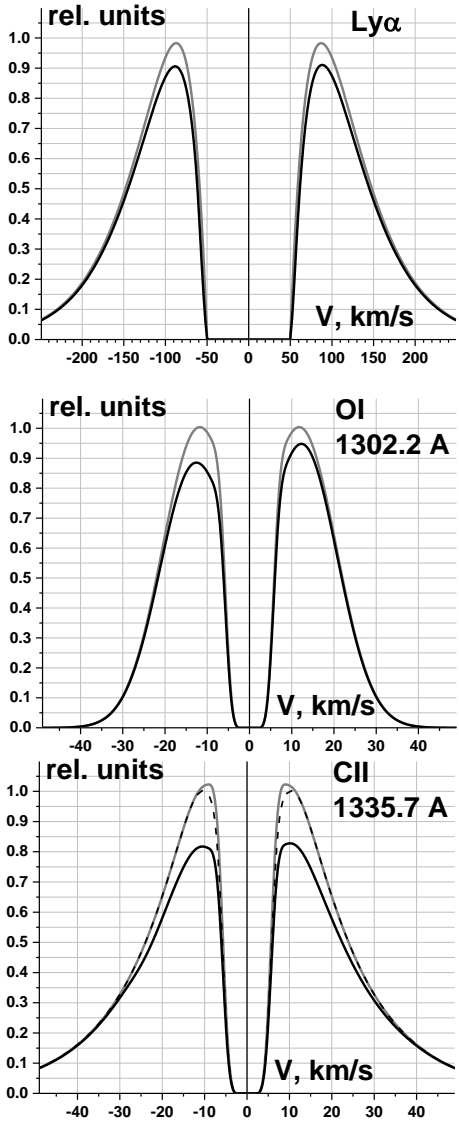

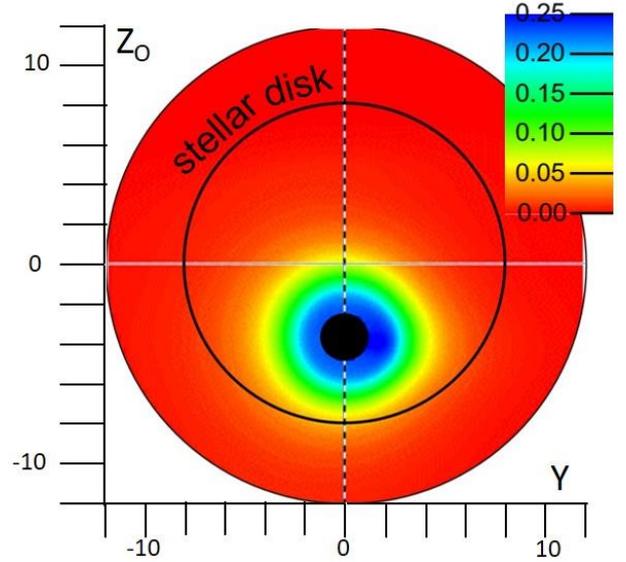

**Figure 7.** Out-of-transit (solid grey) and modeled in-transit (solid black) profiles for HI, OI and CII lines, obtained in the simulation run N1. The center regions of the lines are excluded because of ISM absorption. For CII (bottom panel) also the in-transit line profile produced when the velocity of particles is put to zero is shown (dashed line).

It follows from Figures 6 and 7 that the absorption by OI and CII comes mainly from the region of $|X|<10R_p$. This fact was specifically checked by the calculation of absorption in a range of variable lengths along the LOS. The size of the absorbing region across the LOS is illustrated in Figure 8. The color plot shows the distribution of absorption in CII line, $1-\exp[-\tau(y,z)]$, where $\tau(y,z)$ is the optical depth. As can be seen, most of the CII absorption takes place within the region of $4R_p$ around the planet, which, in spite of the inclination of the planet's orbit, appears during the transit completely inside of the stellar disk. It was specifically checked that there is practically no difference in the absorption between the cases without inclination ($i=90°$) and the real one, $i=86.6°$.

**Figure 8.** The distribution of LOS absorption in CII 1335.7 Å line, as seen by a remote observer at the mid-transit, obtained in simulation run N1. The stellar limb is indicated by black circumference, and the planet by black circle. Note, that this picture uses the observer reference frame, in which the planet is shifted by $\Delta Z=-3.9$ relative the planet-star line ($Z_O=0$, Y=0) due to orbit inclination.

Next we describe absorption by Mg and Si in the corresponding resonant lines. Magnesium atom is rapidly photo-ionized, so the absorption in line of MgI is only slightly larger than the optical depth. At the same time, MgII and SiIII are the most abundant ions of these species at the considered conditions. In Figure 5 one can see that at distance $7R_p$ from the planet their densities are comparable to CII density, though the initial abundances of Mg and Si at the base of the modeled atmosphere were taken an order of magnitude smaller than that of C. In view of the large absorption cross sections of resonant lines of MgII and SiIII one may expect a strong absorption at Doppler shifted velocities <30 km/s. An out-of-transit, and the modeled in-transit SiIII and MgII line profiles are shown in Figure 9. In both lines we have a substantial ~50% absorption which drops to zero at velocities exceeding 40 km/s. An averaged over the whole line, the absorption value for the SiIII is 33%. That significantly exceeds the absorption reported in *Linsky et al. (2010)*. Because the MgI and MgII lines are extremely wide, the averaging gives for the MgII-k line the absorption value of 5.8%. Since no difference for MgII k&h lines has been observed by *Vidal-Majar et al. (2013)* between out and in-transits phases within the measurement accuracy of 2%, the modeled in-transit depth of 5.7% is too large to fit the observations.

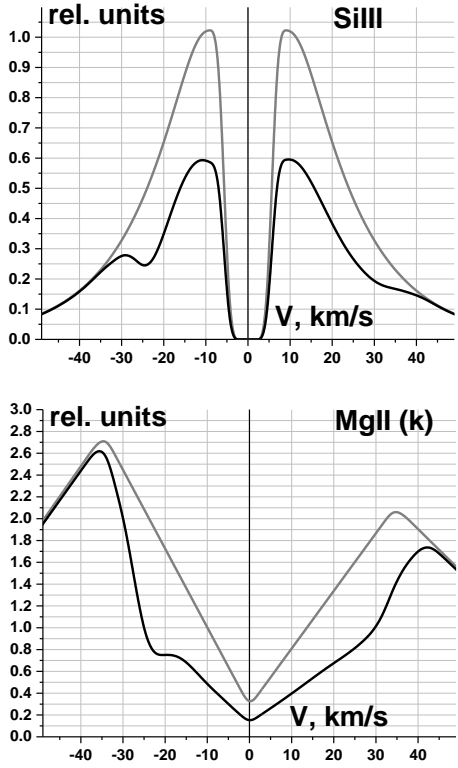

**Figure 9.** An out-of-transit (solid grey) and modeled in-transit (solid black) line profiles for SiIII and MgII resonant transitions, obtained in the simulation run N1.

Next we study the dependence of absorption of various elements on the specific parameters of simulations. In particular, with a series of dedicated simulations we found that the increase of helium abundance in atmosphere results in gradual decrease of the absorption in all considered lines. This is because He increases the mean molecular mass and, thus, decreases the barometric height scale, resulting in faster decrease of all densities with height. Figure 10 demonstrates this effect for HI, OI and MgII lines. The CII and SiIII lines show very small variation with the He/H mixing ratio, and therefore were not displayed. One can see that the result of modeling with an excessive abundance of helium (i.e., He/H>0.2) contradicts the observations, as the obtained absorptions in HI and OI become too small. By the same physical reason, the higher base temperature of atmosphere $T_{base}$, results in higher density and, correspondingly, the larger absorption (modelling runs N1, N8, N9).

Generally speaking, the absorption by elements in the expanding upper atmosphere has to correlate with the overall mass loss rate by the planet, though the degree of this correlation may vary between the elements. Indeed, the planetary mass loss rate is proportional to the density and velocity (i.e. material flux) of the escaping PW, whereas both these quantities increase the absorption. At the same time, the XUV flux, which proportionally increases the mass loss rate, also increases the photo-ionization rate of the absorbing species. Altogether, we see in the modeling results summarized in Table 2 (the runs N1, N5, N6, N7) that the absorption in OI and CII lines decreases with the increasing XUV flux, whereas that in HI line slightly increases and in SiIII line varies in both directions. The strongest effect with the XUV flux is seen for the MgII and HI lines.

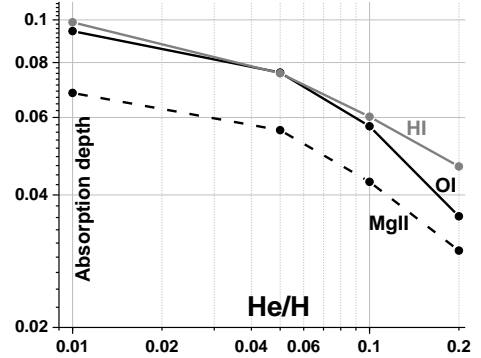

**Figure 10.** The transit depth in HI (gray), OI (black solid) and MgII (black dashed) as a function of helium abundance at the base of the HD209458b atmosphere, obtained in the simulation runs N1, N2, N3, and N4.

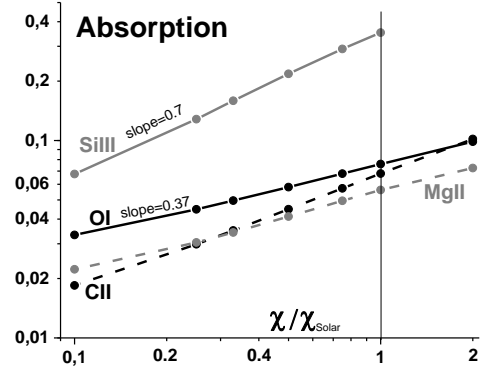

**Figure 11.** The transit depth in OI (black solid), CII (black dashed), SiIII (gray solid), and MgII (gray dashed) lines as a function of the corresponding elements' abundancies at the base of the HD209458b atmosphere, scaled in units of the Solar system values.

Another factor affecting the HD209458b in-transit absorption is analyzed in Figure 11, which shows the dependence of the transit depth in the lines on the corresponding elements' abundancies at the base of the modeled escaping atmosphere, scaled in units of the Solar system values. Important feature following from the simulations is that the absorption by minor elements significantly depends on abundancies, except for MgI line. This is a direct consequence of resonant absorption. Resonant absorption is distributed over wider area than absorption due to the natural line broadening and the related optical depth $\tau$ is relatively small everywhere. In this case the absorption is roughly proportional to the density of species, because $1-e^{-\tau}$

≈τ. Figure 8 confirms that the optical depth outside the planet is indeed significantly less than unity. However, actual dependence is slower than linear, because τ is not very small (τ~0.1), and because at small abundances the absorption by planetary disk 1.5% makes perceptible input.

From Figure 11 it follows that to fit the observations for SiIII and MgII lines at the level of ≤8% and ≤2% respectively we should assume that their abundances are about 10 times less than the Solar. This is a reasonable assumption, taking into account the possible silicate condensation of these elements in the atmosphere of HD209458b.

We also considered the case with the same model parameters as those used in our previous 2D modeling *Shaikhislamov et al. 2018a*. In particular, the primary atmosphere of HD209458b was taken consisting of molecular hydrogen with all the hydrogen chemistry involved, such as $H_2$ dissociation, $H_3^+$ cooling, as well as fixed heating efficiency of 0.5. The results are given in Table 2 as the simulation run N13. Altogether, the mass loss rate of $8 \cdot 10^{10}$ g/s, achieved in this case, is about 3.5 times less, as compared to the standard scenario N1, being at the same time very close to the value obtained in 2D modeling under similar conditions (*Khodachenko et al. 2017, Shaikhislamov et al. 2018a*). The increase of heating efficiency up to unity (the simulation run N14) doubles the mass loss rate and increases absorption in all considered lines. Nevertheless, the absorption values in HI, OI and CII lines obtained with the 3D model at these simulation runs are significantly less than the values yielded by the previous 2D modeling in *Shaikhislamov et al. (2018a)*. Therefore, the more realistic 3D modeling indeed, as was expected, shows significantly weaker tidal acceleration due to Coriolis force spiraling of the PW streams and less absorption in the lines of interest.

In fact, the comparable to observations absorption levels achieved 3D simulations, correspond to 3–4 times higher planetary mass loss rates. In the model without account of hydrogen chemistry and the chemistry of oxygen and carbon based molecules, such mass loss rates of $(2-3) \cdot 10^{11}$ g/s take place at a moderate XUV flux of about 10 erg cm$^{-2}$ s$^{-1}$ at 1 a.u. As it was mentioned above, such a simplifying assumption is supported by the complex aeronomy modeling of *García Muñoz (2007)*, which shows very fast dissociation of $H_2$ in presence of O, C and $H_2O$ even at rather small abundances.

*3.2 Absorption under moderate and strong SW*

In this subsection we consider how the SW plasma flow interacting with escaping PW affects the absorption in lines of the considered minor elements. The simulation run N10 was carried out at the SW values similar to those of the quiet solar wind with the terminal velocity of 400 km/s. The corresponding SW parameters at the planet orbit are: $V_{sw}$=220 km/s, $T_{sw}$=0.8 MK, $n_{sw}$=10$^4$ cm$^{-3}$, $p_{sw}$=10$^{-5}$ µbar. The escape regime realized in this case is close to one, we called as 'captured by the star' in our previous 2D simulations (*Shaikhislamov et al. 2016, 2018a, Khodachenko et al. 2017*). As the results show, under such moderate SW conditions the absorption in considered lines remains practically the same, as in the case of a very weak SW. The influence of SW plasma flow on the structure and shape of the escaping PW streams becomes pronounced relatively far from the planet at distances beyond ~40$R_p$, where SW deflects the star-directed PW steam and pushes away the tail stream, preventing accumulation of the escaping planetary material around the star.

Increasing the SW pressure by 3 times (run N11) still doesn't significantly change the absorption by elements. However, for an order of magnitude higher total pressure of SW (the simulation run N12), achieved for example in Coronal Mass Ejections, the 'blown by the wind' regime of PW and SW interaction is realized, in which the star-directed PW stream ahead of the planet is stopped by the SW and redirected into the tail. The structure of interacting PW and SW flows under the conditions of moderate (run N10) and strong (run N12) SW is demonstrated in Figure 12. Similar to our previous 2D and 3D simulations (*Khodachenko et al. 2017, 2019, Shaikhislamov et al. 2018a,b*), we observe here the generation of large number of Energetic Neutral Atoms (ENAs) in the shocked region between the bowshock and the ionopause. The equatorial and meridional cuts of the ENAs distribution and electron temperature in the shocked region are plotted in Figure 13.

The ENAs not only increase the Lyα in-transit absorption up to almost 20%, but also qualitatively change its character. As it can be seen in Figure 14, the Lyα absorption in the case of strong SW is essentially asymmetric and covers the range of blue-shifted velocities [–50; –200] km/s, indicating on the presence of fast hydrogen atoms moving away from the star. Such scenario is in agreement with the interpretation of observations, proposed in *Vidal-Majar (2003, 2004, 2008)*. Note, however, that while the absorption in HI is significantly increased under the strong SW conditions, the absorption in lines of minor species behaves oppositely. The decrease of absorption in lines of OI, CII, SiIII and MgII is explained by the fact that the strong SW reduces the region ahead of the planet where the escaping PW material can be accelerated beyond velocities of 20 km/s, to enable efficient detectable absorption in these lines. This explanation is supported by the line profiles, which show significantly stronger absorption in the blue wings and rather small absorption in the red wings. In Figure 14 the asymmetry effect is shown for HI (Lyα) and CII lines. Therefore, the in-transit measurement of absorption lines of minor species, if spectrally resolved, may show an asymmetry caused by strong SW.

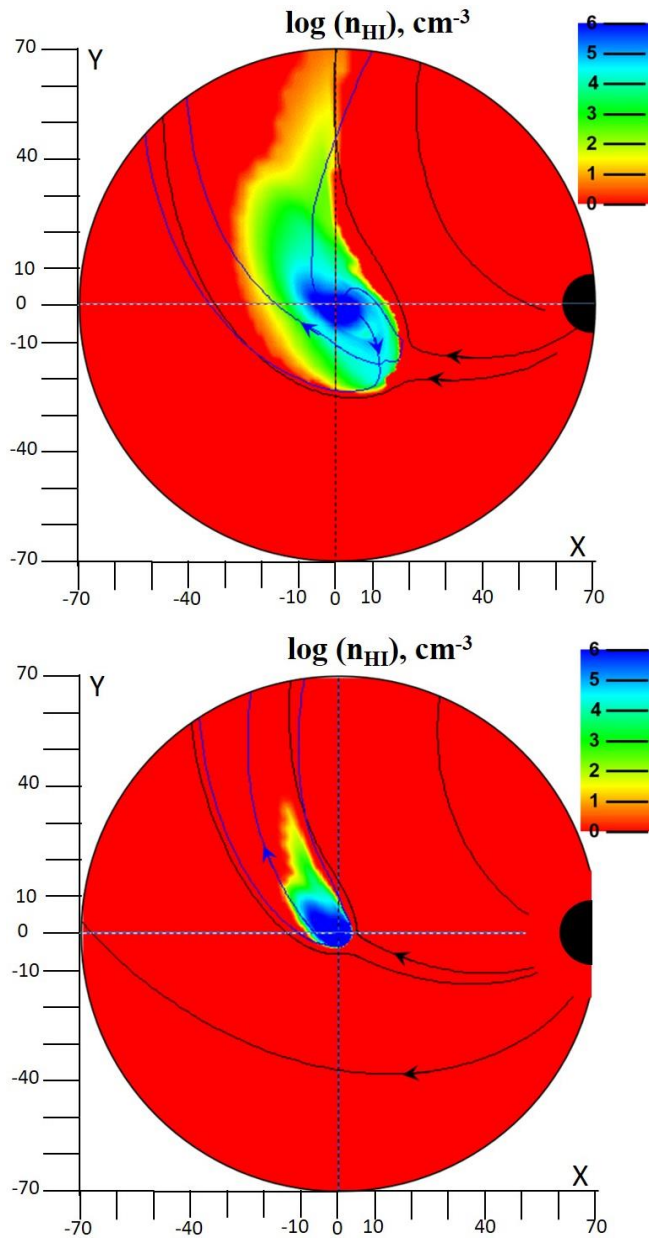

**Figure 12.** Atomic hydrogen density distributions ($n_{HI}$, in log-scale), obtained under conditions of a moderate (*upper panel*; the simulation run N10) and strong (*bottom panel*; the simulation run N12) SW. The flows of stellar protons and planetary hydrogen atoms are shown with black and blue streamlines, respectively.

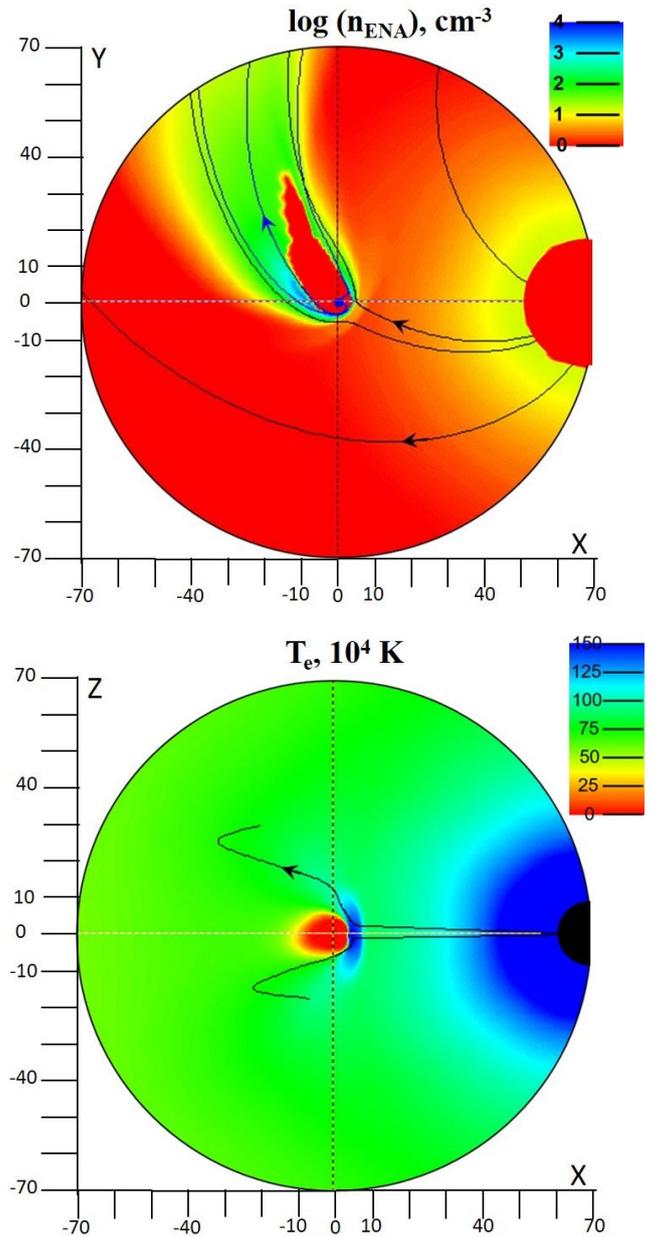

**Figure 13.** Distributions of ENA density (*top panel*; log-scale, equatorial plane) and electron temperature (*bottom panel*; meridional plane), under the conditions of a strong SW, obtained in the simulation N12. The flows of stellar protons and planetary hydrogen atoms are shown with black and blue streamlines, respectively.

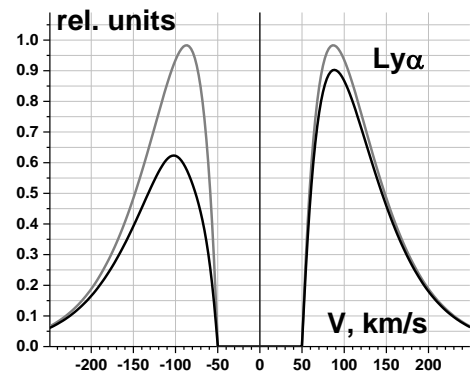

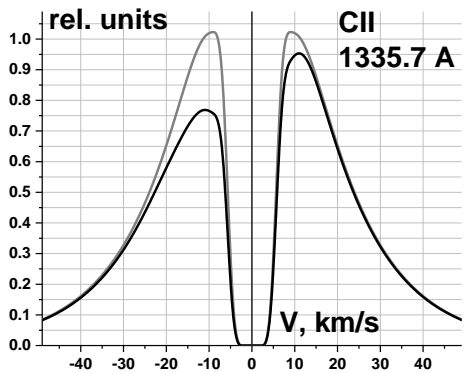

**Figure 14.** Out-of-transit (solid grey) and modeled in-transit (solid black) absorption profiles for HI (*upper panel*) and CII (*bottom panel*) lines, obtained under strong SW conditions in the simulation run N12.

## 4 DISCUSSION AND CONCLUSION

Using the global 3D multi-fluid hydrodynamic model, we simulated the in-transit absorption depths in all the FUV lines, observed for HD209458b, and compared the new results with the outcomes of our previous similar study, which was based on 2D model. For the first time, the absorption by different elements was simulated on the basis of the self-consistent approach which describes the heating and ionization of the multi-component upper planetary atmosphere by the absorbed stellar XUV, the atmospheric material photo-chemistry and hydrodynamic escape beyond the Roche lobe where it interacts with SW on the global 3D scale of the whole stellar-planetary system.

So far, the 3D MHD models applied for similar studies of hot close-orbit giant exoplanets were neither multi-fluid nor calculating the absorption by minor elements (e.g., *Bisikalo et al. 2013, 2917, Matsakos et al. 2015, Tripathi et al. 2015*). Moreover, all of them simulated the escaping PW starting from a prescribed layer in the upper planetary atmosphere with an empirically chosen maximum temperature, or considered the material flow propagating already beyond the Roche lobe surface. In fact, as shown in Figure 4, the temperature maximum in the expanding upper atmosphere of a hot Jupiter may be as far as $2R_p$. The simplified non-self-consistent approach with a prescribed maximum temperature at the inner boundary of the simulation domain not only a-priori excludes from the consideration the region, which is most important for the absorption by HI and MgI, but also makes impossible correct study of the role of the varying XUV flux and other parameters of the modeled system.

Among the goals of the present study was the validation of our previous results obtained with a 2D model (*Shaikhislamov et al. 2018a*). Because the most of the absorption in OI, CII, SiIII and MgII lines comes from the PW flow beyond the Roche lobe, where it is accelerated by stellar gravity, the crucial question was to which extend the spiraling of the escaping PW stream and its actually achieved velocity may influence (e.g., support or alter) the conclusions made on the basis of 2D modeling. An answer obtained in the present work is that the PW velocity is smaller than one calculated in 2D simulation, but it is still sufficient to cover the widths (±40 km/s) of the lines of interest and to produce the absorption comparable to observations.

The main conclusions made on the basis of 2D modeling have been verified and elaborated further with the 3D simulations. Specifically, the in-transit absorption in HI, OI and CII resonant lines at the level of (7–10) % and very deep absorption in MgII and SiIII lines were confirmed, if the typical Solar system abundancies for these elements are assumed; an insignificant influence of the moderate SW on the absorption values has been verified, whereas strong SW was shown to result in red-blue asymmetry of absorption profiles.

The model used in the present paper is built with a significantly improved approach to the calculation of ionization stages of the minor elements of interest and the heating efficiency by the absorbed XUV. It also shows that the spiraling of PW streams due to the Coriolis force declining the material flow from the planet-star direction, causes a lower absorption levels than those predicted by the 2D model. One of the important findings is that the comparable to observations absorption levels above 7% in the lines of HI, OI and CII are achieved at mass loss rate of the HD209458b of about $(2\text{--}3)\cdot 10^{11}$ g/s. This is 3.5 times higher than the values obtained in our previous modeling with the complete hydrogen chemistry involved. Such mass loss rate is reached at a moderate level of XUV flux assuming fast dissociation of molecular hydrogen in the upper atmosphere of HD209458b. This assumption is supported by the complex 1D aeronomy modeling by *García Muñoz (2007)*, who has shown that at the Solar system abundancies of oxygen, the destruction of molecular hydrogen in atmosphere proceeds via reaction $O+H_2 \rightarrow OH+H$, which reduces the height of the $H_2$ half-dissociation level below $0.1R_p$ and the corresponding pressures above $10^{-6}$ bar. Therefore, the used simplification of omitting the molecular chemistry appears a relevant approximation for the considered problem.

The telling difference between previously applied 1D models to calculated absorption in observed lines (e.g., *Ben-Jaffel & Sona Hosseini 2010, Koskinen et al. 2013b*) and 3D model is that to match observations the former required significantly higher than the Solar system abundances, while the latter predicts the Solar or even lower abundances. This difference is connected with the acceleration of escaping PW flow by the stellar gravity beyond the Roche lobe, which cannot be properly treated in 1D.

Another important finding of the 3D modeling concerns the abundances of Mg and Si. The non-detection of absorption by MgII-k&h lines within the measurement accuracy of 2% puts a direct constrain on the simulation results. The same is true for the in-transit measurements of SiIII line by *Linsky et al. 2010*. While the stellar variability in these measurements can influence the true absorption level, the upper limit of <8% is also a direct constrain. Because the MgII and SiIII are most populous ion stages of these elements and resonant transitions have large cross-sections, the low absorption level requires explanation. In the present work to fit these constrains, we have to take abundances of both of Mg and Si by 10 times less than the Solar values. The lower rather than higher abundances of these elements, as compared to the Solar system values, are in full accord with the fact that their condensation points are definitely reached in the lower atmosphere of HD209458b.

The revealed influence of the helium abundance on the HI, OI and MgII absorption (see also *Shaikhislamov et al. 2018b*) and the obtained constrains in that respect, may be of interest regarding the recent observations in He lines (*Oklopčić et al. 2018*). In particular, it has been shown that the absorption in all considered lines decreases at He abundancies above 0.1 due to the decrease of atmospheric height scale and the related steeper decrease of the density with the height.

The future FUV observations being combined with the presented 3D modeling may put stricter constrains on the abundances of elements, as well as on metallicity and C/O ratio. Spectrally resolved measurements of the absorption lines of minor elements, especially in combination with Lyα measurements, may provide a first consistent prove of existence of strong stellar plasma winds, whereas the simulation of measured values would enable its quantitative probing.

**Acknowledgements:**
This work was supported by grant № 18-12-00080 of the Russian Science Foundation. HL and MLK acknowledge the Austrian Science Fund (FWF) projects S11606-N16, S11607-N16 and I2939-N27 of the Austrian Science Foundation (FWF). HL acknowledge also support from the FWF project P25256-N27 ('Characterizing Stellar and Exoplanetary Environments via Modeling of Lyman-α Transit Observations of Hot Jupiters'). IS acknowledges productive discussions during the Programme "Astrophysical Origins: Pathways from Star Formation to Habitable Planets" at ESI, Wien, Austria in July 2019. Parallel computing simulations, key for this study, have been performed at Computation Center of Novosibirsk State University, SB RAS Siberian Supercomputer Center, Joint Supercomputer Center of RAS and Supercomputing Center of the Lomonosov Moscow State University.